\documentstyle[12pt,a4,psfig]{article}
\def\eg{{\it e.g.}}
\def\chg{\tilde{\omega}}
\def\cosd{\cos\delta}
\def\kplusdecay{K^+ \to \pi^+ \nu \nubar}

\def\ie{{\it i.e.}}
\def\gev{{\rm GeV} }

\def\vsk#1{\noalign{\vskip#1 cm}}
\def\vsp#1{\vspace{#1 cm}}
\def\ov{\overline}
\def\xd{x_d^{}}
\def\ek{\epsilon_K^{}}

\def\xj{x_j}
\def\zi{z_i}
\def\ztild{\tilde{z}}
\def\sa{s_\alpha^{}}
\def\sb{s_\beta^{}}

\def\rim{r_{im}^{}}
\def\rin{r_{in}^{}}
\def\tlk{t_{\ell k}^{}}
\def\ra{r_\alpha^{}}
\def\rb{r_\beta^{}}
\def\si{s_i^{}} 
\def\sj{s_j^{}} 
\def\sn2w{\sin^2\theta_W}
\def\exi{x_i}
\def\disp{\displaystyle}

\def\lsim{{\mathop <\limits_\sim}}

\def\gm5{\gamma_5}

\def\xt{x_t^{}}
\def\xh{x_H^{}}

\def\to{\rightarrow} 
\def\longto{\longrightarrow} 
\def\nubar{\ov{\nu}}
\def\bbbar{B^0\mbox{-}\ov{B^0}}
\def\kkbar{K^0\mbox{-}\ov{K^0}}
\def\etal{{\it et al.}}
\newcommand{\beq}{\begin{equation}}
\newcommand{\eeq}{\end{equation}}
\newcommand{\bea}{\begin{eqnarray}}
\newcommand{\eea}{\end{eqnarray}}
\newcommand{\bsub}{\begin{subequations}}
\newcommand{\esub}{\end{subequations}}

\renewcommand{\theequation}{\thesection.\arabic{equation}}
\newcommand{\clean}{\setcounter{equation}{0}}

\def\PRD#1#2#3{Phys. Rev. {\bf D#1} (19#2) #3}
\def\NPB#1#2#3{Nucl. Phys. {\bf B#1} (19#2) #3}
\def\ZPC#1#2#3{Z. Phys. {\bf C#1} (19#2) #3}
\def\PLB#1#2#3{Phys. Lett. {\bf B#1} (19#2) #3}
\def\PRL#1#2#3{Phys. Rev. Lett. {\bf #1} (19#2) #3}
\def\PRep#1#2#3{Phys. Rep. {\bf #1} (19#2) #3}
\def\PTP#1#2#3{Prog. Theor. Phys. {\bf #1} (19#2) #3}
\def\MPL#1#2#3{Mod. Phys. Lett. {\bf A#1} (19#2) #3}

\def\figmssm{
\begin{figure}[t]
\begin{center}
\leavevmode\psfig{figure=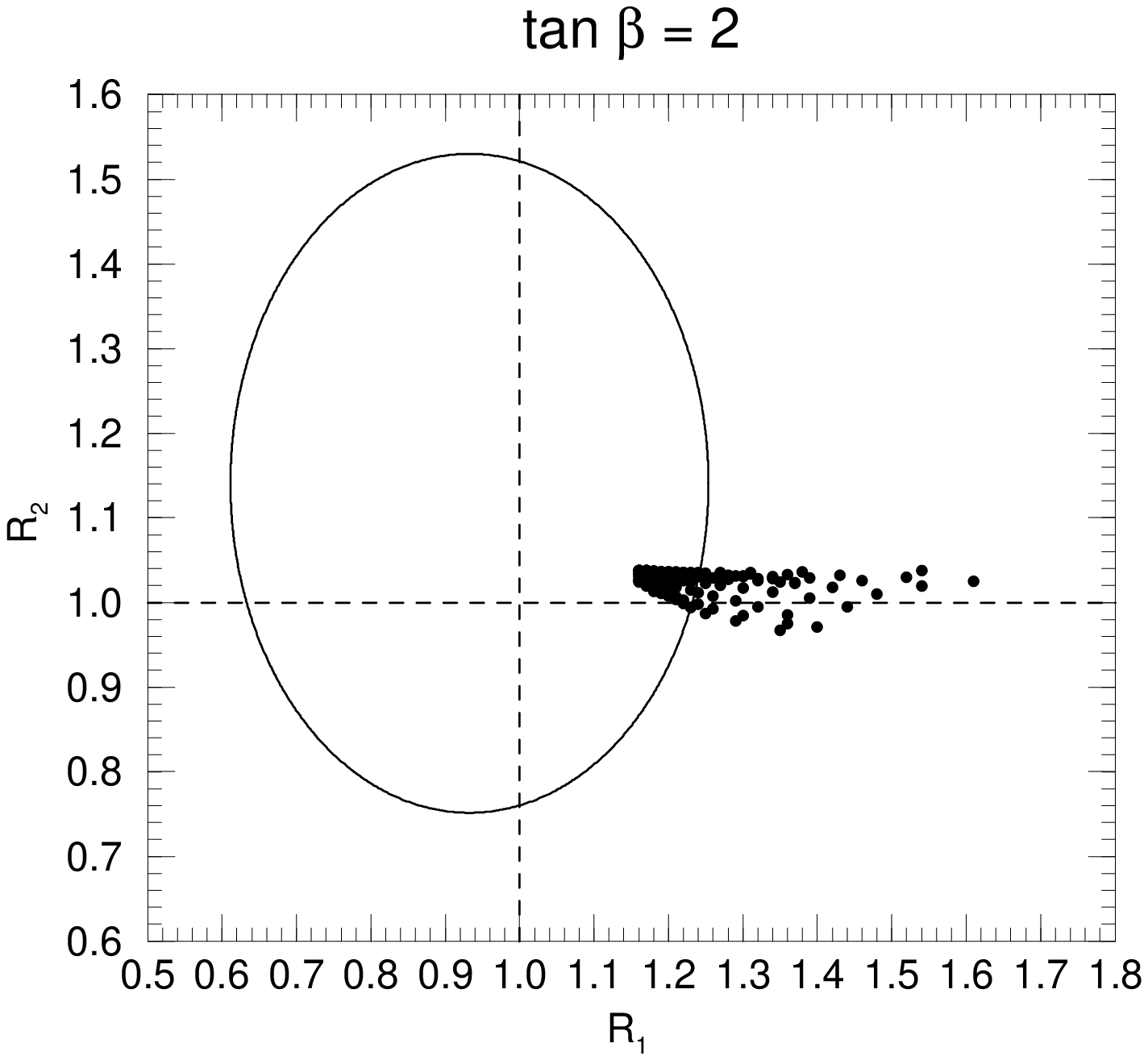,width=6cm}
\leavevmode\psfig{figure=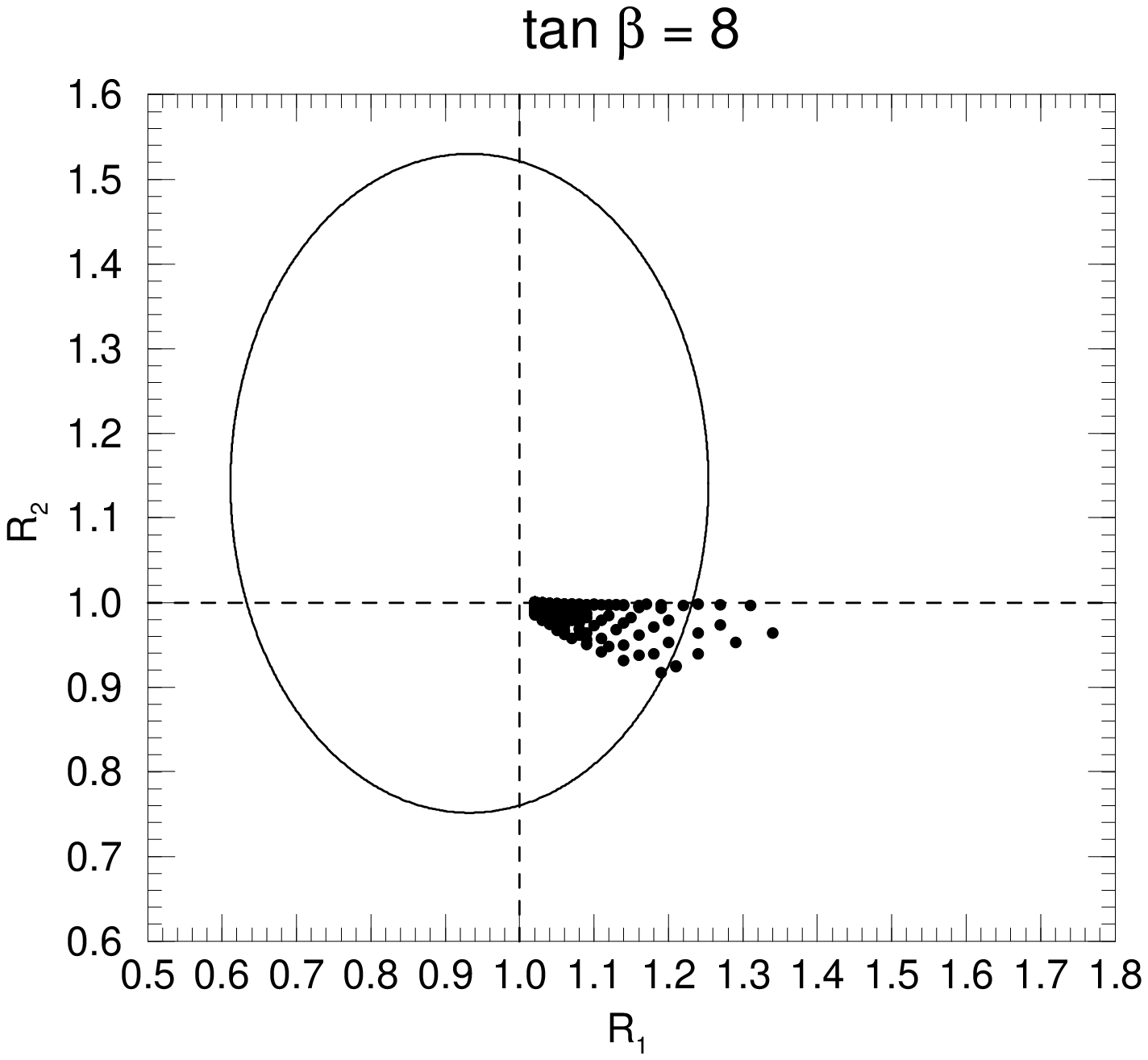,width=6cm}
\end{center}
\caption{
The MSSM contributions to $R_1,R_2$ parameters for 
$\tan\beta = 2$ (left) and $\tan\beta = 8$ (right). 
The 1-$\sigma$ allowed region of $R_1,R_2$ parameters 
for $\cosd = 0.36$ is also shown. 
}
\label{r2_mssm}
\vsp{0.5}
\end{figure}
}
\def\figthdm{
\begin{figure}[t]
\begin{center}
\leavevmode\psfig{figure=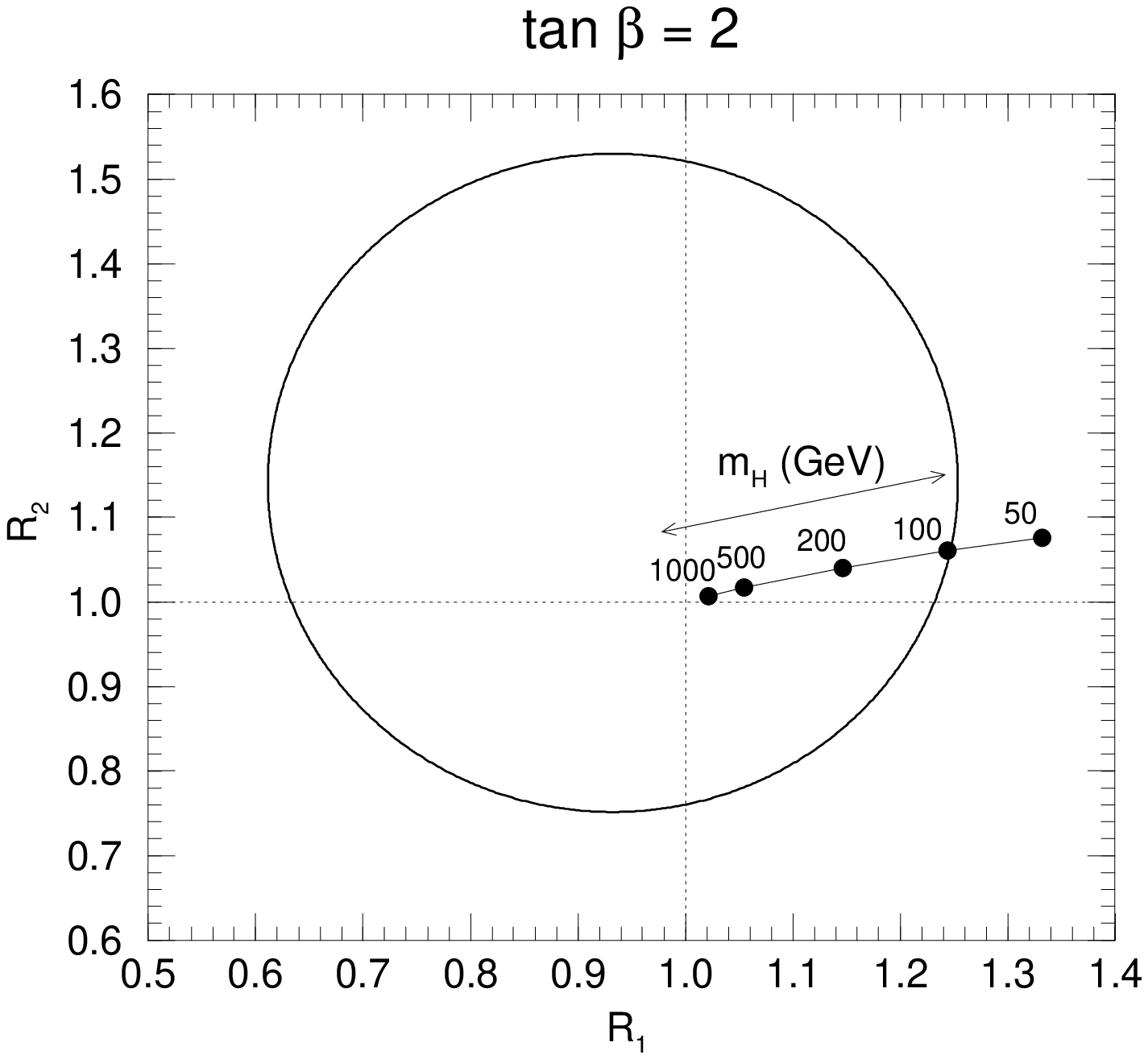,width=9cm}
\end{center}
\caption{ 
The THDM contributions to 
$R_1,R_2$ parameters for $\tan\beta=2$. 
The 1-$\sigma$ allowed region of $R_1,R_2$ parameters 
for $\cosd = 0.36$ is also shown. 
}
\label{r2_thdm}
\vsp{0.5}
\end{figure}
}
\def\figbbkkonesigma{
\begin{figure}[t]
\begin{center}
\leavevmode\psfig{figure=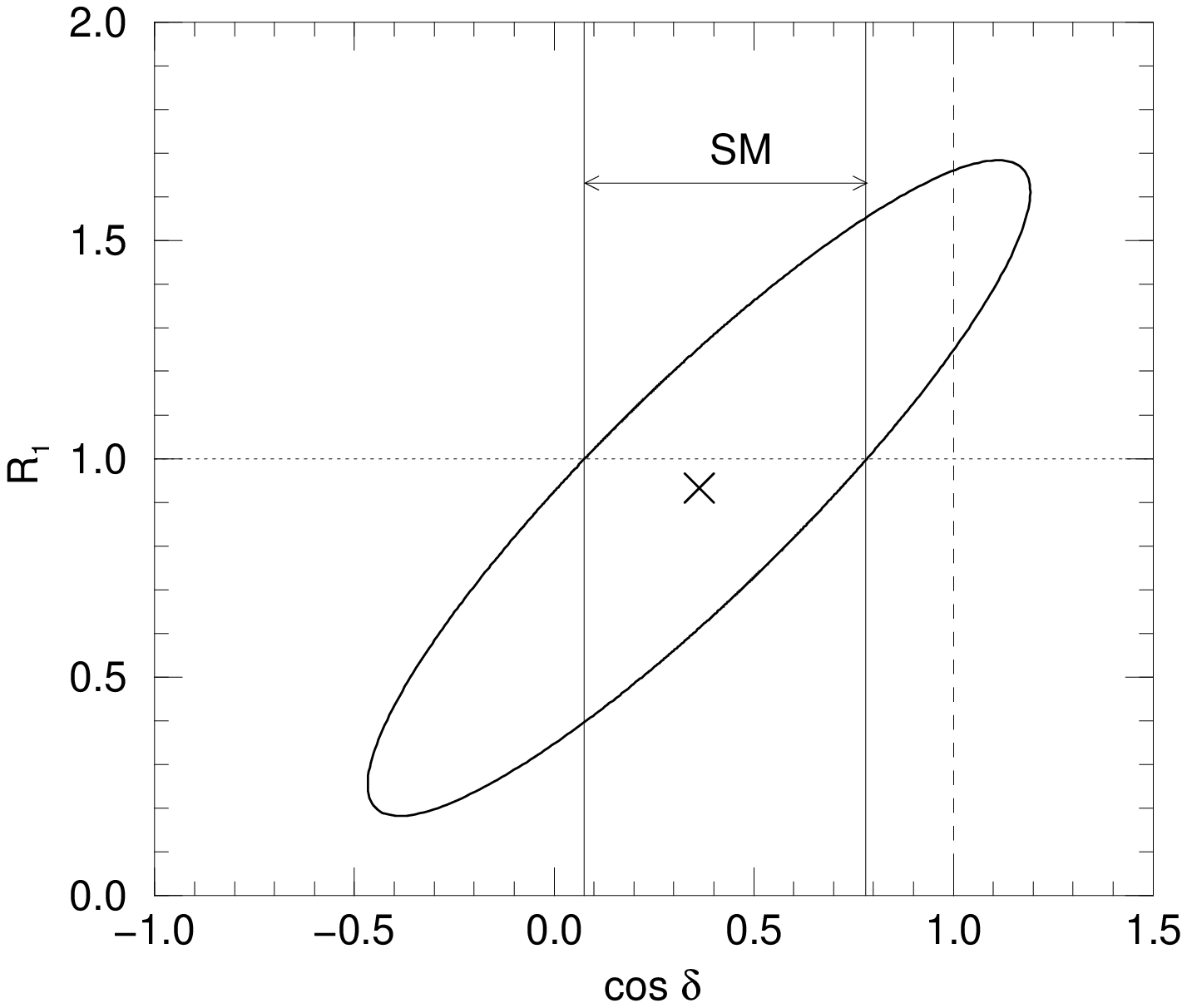,width=9cm}
\end{center}
\caption{ 
The 1-$\sigma$ (39\% CL) allowed region from the experimental 
results of the $\bbbar$, $\kkbar$ mixings. 
The range between the two solid lines is the allowed region 
of $\cosd$ in the SM. 
}
\label{bbkk39cl}
\vsp{0.5}
\end{figure}
}
\def\figsumr1r2{
\begin{figure}[t]
\begin{center}
\leavevmode\psfig{figure=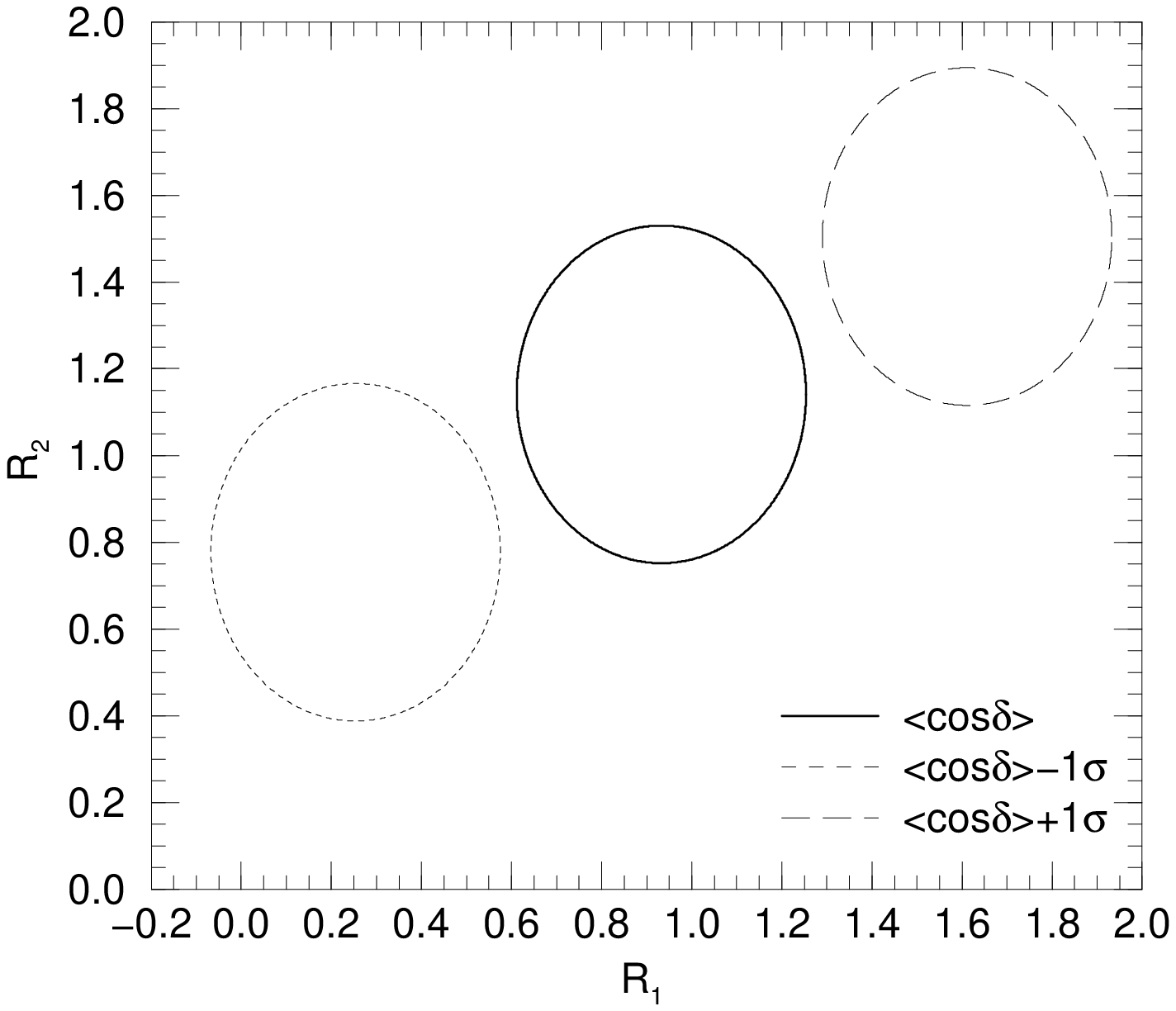,width=9cm}
\end{center}
\caption{The 1-$\sigma$ allowed regions of $R_1,R_2$ 
parameters. Three contours are corresponding to 
$\cosd = 0.36$ (solid line), 
$\cosd = -0.47$ (dotted line) and 
$\cosd = 1.19$ (dashed line), respectively. }
\label{sum_r1r2}
\vsp{0.5}
\end{figure}
}
\makeatletter
\newtoks\@stequation

\def\subequations{\refstepcounter{equation}%
  \edef\@savedequation{\the\c@equation}%
  \@stequation=\expandafter{\theequation}
  \edef\@savedtheequation{\the\@stequation}
  \edef\oldtheequation{\theequation}%
  \setcounter{equation}{0}%
  \def\theequation{\oldtheequation\alph{equation}}}

\def\endsubequations{%
  \ifnum\c@equation < 2 \@warning{Only \the\c@equation\space subequation
    used in equation \@savedequation}\fi
  \setcounter{equation}{\@savedequation}%
  \@stequation=\expandafter{\@savedtheequation}%
  \edef\theequation{\the\@stequation}%
  \global\@ignoretrue}

\def\eqnarray{\stepcounter{equation}\let\@currentlabel\theequation
\global\@eqnswtrue\m@th
\global\@eqcnt\z@\tabskip\@centering\let\\\@eqncr
$$\halign to\displaywidth\bgroup\@eqnsel\hskip\@centering
     $\displaystyle\tabskip\z@{##}$&\global\@eqcnt\@ne
      \hfil$\;{##}\;$\hfil
     &\global\@eqcnt\tw@ $\displaystyle\tabskip\z@{##}$\hfil
   \tabskip\@centering&\llap{##}\tabskip\z@\cr}

\makeatother

\begin{document}
\vspace*{-15mm}
\baselineskip 10pt
\begin{flushright}
\begin{tabular}{l}
{\bf KEK-TH-554}\\
{\bf hep-ph/9801406}\\
{\bf January 1998}
\end{tabular}
\end{flushright}
\baselineskip 18pt 
\vglue 15mm 
\begin{center}
\baselineskip 24pt
{\Large\bf
Impacts on searching for signatures of new physics 
from $\kplusdecay$ decay
}
\vspace{5mm}

\def\thefootnote{\alph{footnote}}
\setcounter{footnote}{0}
{\bf
Gi-Chol Cho\footnote{Research Fellow of the Japan Society 
for the Promotion of Science}
}

\vspace{5mm}
\def\thefootnote{\arabic{footnote}}
\setcounter{footnote}{0}
{\it Theory Group, KEK, Tsukuba, Ibaraki 305, Japan}
\vspace{20mm}
\end{center}
\begin{center}
{\bf Abstract}\\[10mm]
\begin{minipage}{12cm}
\noindent
\baselineskip 18pt
We study impacts on new physics search from the rare decay 
$\kplusdecay$. 
In a certain class of new physics models, the extra 
contributions to FCNC processes can be parametrized by its 
ratio to the standard model (SM) contribution with the common 
CKM factors. 
The ratio $R_1$ has been used in the analysis of 
$\xd$ and $\ek$ parameters. 
In the above class of models, the $\kplusdecay$ decay 
amplitude can be parametrized by the ratio $R_2$.  
Then the experimentally allowed region for new physics 
contributions can be given in terms of $R_1,R_2$ and the CP 
violating phase $\delta$ of the CKM matrix. 
Constraints on $R_1$ and $\cosd$ are obtained by taking 
account of current experimental data and theoretical 
uncertainties on $\bbbar$ and $\kkbar$ mixings. 
We study impacts of future improved measurements 
by using $(R_1, R_2, \cosd)$ basis. 
We discuss contributions in the minimal supersymmetric 
SM and the two Higgs doublet model as examples. 
\end{minipage}
\end{center}
\newpage
\baselineskip 18pt 
\section{ Introduction }
Processes mediated by flavor changing neutral current (FCNC) 
have been considered as good probes of physics beyond the 
standard model (SM). 
By using the experimentally well measured processes, 
an existence of new physics may arise as violation of 
the unitarity of the Cabbibo-Kobayashi-Maskawa (CKM) matrix.  
Such signatures of new physics will be explored through 
the determination of the unitarity triangle 
at B-factories at KEK and SLAC in the near future. 

The rare decay $K^+ \to \pi^+ \nu \nubar$ is one of the most 
promising processes to extract clean informations about 
the CKM matrix elements~\cite{buchalla-buras}.
The decay rate has small theoretical uncertainties 
because the interactions are dominated by the short-distance 
physics.  
The long-distance contributions have been estimated as 
$10^{-3}$ smaller than the short-distance 
contributions~\cite{long_distance}. 
The importance of this decay mode on the determination 
of the unitarity triangle has been discussed in 
\cite{buras,buchalla2}. 
Furthermore, 
by carefully examining the consistencies of the 
CKM matrix elements measured from this decay process and the 
other processes, we may find a signature of new physics. 

Recently, E787 collaboration reported the first observation 
of an event consistent with this decay process, and obtained  
${\rm Br} (K^+ \to \pi^+ \nu \nubar) 
= 4.2^{+9.7}_{-3.5} \times 10^{-10}$~\cite{E787}.   
Although there is still only one candidate event, 
the report motivates us to examine the implication of 
the above estimate of the branching fraction and of 
its improvement in the near future. 

In this paper, we study impacts on the search for a new physics 
signal from the $K^+ \to \pi^+ \nu \nubar$ decay in a class of 
new physics models that satisfy the following two conditions:
(i) The flavor mixing in the new physics sector is governed by 
the SM CKM matrix elements, 
(ii) The main contributions to the FCNC processes are given 
through loop effects mediated by the third generation particles. 
The condition (i) means that the effective Lagrangian 
of the FCNC processes in the new physics sector 
can be described by the same form with that of 
the SM besides the estimation of the loop contributions. 
The condition (ii) implies either 
the extra contributions from the first two generations 
do not differ so much, or those are negligible as compared 
with the contribution from the third generation. 

Our assumptions can be valid not only in 
$K^+ \to \pi^+ \nu \nubar$ decay but also in 
other FCNC processes, such as $\bbbar$ and $\kkbar$ mixings. 
We will show that new physics contributions to those 
processes can be parametrized by two quantities, $R_1$ for 
$\bbbar,\kkbar$ mixings and $R_2$ for $\kplusdecay$ decay. 
Both quantities are defined as the ratio of the new physics 
contribution to that of the SM. 
Taking account of current experimental data on 
$\xd$ and $\ek$ parameters in $\bbbar$ and $\kkbar$ mixings,  
and uncertainties in the hadronic parameters, 
we show constraints on the new physics contributions 
in terms of $R_1$ and $\cosd$, where $\delta$ is the CP 
violating phase of the CKM matrix in the standard 
parametrization~\cite{PDG}. 
We also find constraints on $R_1$, $R_2$ and $\cosd$ 
by assuming the future improvement of the 
Br($\kplusdecay$) measurements. 
As examples of new physics models which naturally satisfy 
the conditions (i) and (ii), 
we examine the consequences of the minimal supersymmetric 
standard model (MSSM)~\cite{SUGRA} and the two Higgs doublet 
model (THDM)~\cite{higgs_hunter}.
\section{New physics contributions to the FCNC processes 
	in the $B$ and $K$ meson systems}
\clean
The effective Lagrangian for the $\kplusdecay$ process in the SM 
is given by~\cite{inami-lim}: 
\beq
{\cal L}^{K^+}_{eff} = \frac{G_F}{\sqrt{2}} \frac{2\alpha(m_Z)}{\pi} 
	\frac{1}{\sin^2 \theta_W} 
	\ov{\nu}_\ell \gamma^\mu P_L \nu_\ell\ 
	\ov{s}\gamma_\mu P_L d\ 
	\sum_{i = 2,3} V_{i2}^* V_{i1}^{}\ 
	\eta_i D_W^{}(i),
\label{lagrangian_kdecay}
\eeq
where $i$ and $\ell$ are the generation indices for the 
up-type quarks and leptons, respectively. 
The CKM matrix element is given by $V_{ij}$ and 
the projection operator $P_L$ is defined as 
$P_L \equiv (1-\gamma_5^{})/2$. 
The QCD correction factor and the loop function are denoted 
by $\eta_i$ and $D_W(i)$, respectively. 
The top quark loop function is given as~\cite{inami-lim}: 
\bea
D_W^{}(3) &=& \frac{\xt}{8}\biggl\{ 
\frac{\xt + 2}{\xt-1} + \frac{3\xt-6}{(\xt-1)^2}\ln \xt 
\biggr\}, 
\eea
where $\xt = m_t^2/m_W^2$. 
The corresponding QCD correction factor has been estimated as 
$\eta_3 = 0.985$ for $170~{\rm GeV} \le m_t^{} \le 
190~{\rm GeV}$~\cite{buchalla}.
The charm quark loop function with the QCD 
correction is numerically given as $ \eta_2^{} D_W(2) = 
\lambda^4\times(0.40 \pm 0.06 )$~\cite{buchalla2} 
where $\lambda \equiv |V_{12}|$.
The error is due to uncertainties in 
the charm quark mass and higher order QCD corrections. 
Then, summing up the three generations of neutrino, 
the branching ratio is expressed as~\cite{marciano}
\bea
{\rm Br}(K^+ \rightarrow \pi^+ \nu \ov{\nu}) &=& 
3\times 
\frac{G_F^2}{192\pi^3} \biggl( \frac{\alpha(m_Z)}{2\pi \sn2w} 
\biggr)^2 \biggl | f_+^{K^+ \pi^+}(0) \biggr |^2 
I(m_{K^+}, m_{\pi^+}) \tau_{K^+} 
\nonumber \\
&& \times 
\biggl | V_{32}^* V_{31}^{} \eta_3^{} D_W^{}(3) 
	+ V_{22}^* V_{21}^{}  \eta_2^{} D_W^{}(2)\biggr|^2  
\nonumber \\
&=& 1.57 \times 10^{-4}  
\biggl | V_{32}^* V_{31}^{} \eta_3^{} D_W^{}(3) 
	+ V_{22}^* V_{21}^{}  \eta_2^{} D_W^{}(2)\biggr|^2,   
\label{br_kdecay}
\eea
where $\tau_{K^+}$ denotes the lifetime of the $K^+$ meson. 
The function $I(m_{K^+}, m_{\pi^+})$ gives the phase space 
factor 
and the form factor $f_+^{K^+ \pi^+}(0)$ contains 
the SU(3)-breaking quark mass effects. The explicit form of 
$I(m_{K^+}, m_{\pi^+})$ can be found in \cite{marciano}.
With the above estimates for the loop functions and the QCD 
correction factors, the branching ratio 
is predicted to be~\cite{recent_estimation}
\beq
{\rm Br}(K^+ \rightarrow \pi^+ \nu \ov{\nu})_{\rm SM} = 
(9.1 \pm 3.8)\times 10^{-11}, 
\eeq
in the SM, where the error is dominated by the 
uncertainties of the CKM matrix elements. 

The effective Lagrangian of the $\bbbar$ mixing in the SM 
is expressed by 
\bea
{\cal L}^{\Delta B =2}_{eff} 
	= \frac{G_F^2 M_W^2}{4\pi^2}~ 
	\ov{d}\gamma^\mu P_L b~ \ov{d}\gamma_\mu P_L b 
	\sum_{i,j=2,3} 
	V_{i1}^* V_{i3}^{} V_{j1}^* V_{j3}^{}~ 
	F_V^W(i,j). 
\label{lagrangian_bbmixing}
\eea
Likewise, ${\cal L}^{\Delta S =2}_{eff}$  
for the $\kkbar$ mixing is obtained by replacing $V_{i3}$ 
with $V_{i2}$, and the $b$-quark operators  
with the $s$-quark ones, respectively. 
The loop function $F_V^W(i,j)$ is given by~\cite{inami-lim}
\bsub
\bea
F_V^W(i,j) &=& -\frac{1}{4} \exi \xj \biggl\{ 
\frac{\exi^2 - 8 \exi + 4}{(\exi - \xj)(\exi -1)^2} \ln \exi 
+ 
\frac{\xj^2 - 8 \xj + 4}{(\xj - \exi)(\xj -1)^2} \ln \xj 
\nonumber \\
&&~~ 
- \frac{3}{(\exi - 1)(\xj - 1)}\biggr\}, \\
\vsk{0.3}
F_V^W(i,i) &=& -\frac{1}{4} \biggl(\frac{\exi}{\exi-1}\biggr)^2 
\bigl\{ \exi - 11 + \frac{4}{\exi} + \frac{6\exi}{\exi-1}\ln \exi 
\bigr\}, 
\eea
\esub
where $\exi$ is defined by $\exi \equiv m_{u_i}^2/m_W^2$. 
The $B$-meson mixing parameter $\xd$ is defined by 
$\xd \equiv \Delta M_B/ \Gamma_B$, where $\Delta M_B$ and 
$\Gamma_B$ correspond to the $B$-meson mass difference and the 
average width of the mass eigenstates, respectively. 
The mass difference is induced by the above $\Delta B=2$ 
operator~(\ref{lagrangian_bbmixing}) and we can express 
the mixing parameter $\xd$ in the SM as 
\beq
\xd = \frac{G_F^2}{6\pi^2}M_W^2 \frac{M_B}{\Gamma_B} f_B^2 B_B 
	\bigl| V_{31}^* V_{33}^{} \bigr|^2 \eta_B 
	\bigl| F_V^W (3,3) \bigr|, 
\label{xd_bb}
\eeq
where $f_B, B_B$ and $\eta_B$ denote the decay constant of 
$B^0$-meson, the bag parameter of $\bbbar$ 
mixing and the short-distance QCD correction factor, respectively.

For the $\kkbar$ system, it is known that the theoretical 
prediction for the mass difference $\Delta M_K$ cannot be given 
precisely because it receives the large long-distance contributions. 
On the other hand, the CP-violating parameter $\ek$ is 
dominated by the short-distance contributions which are 
given by the imaginary part of the same box diagram 
of the $\bbbar$ transition besides the external quark lines. 
We can express the $\ek$ parameter in the SM as 
\bea
\ek &=& -e^{i\pi/4} \frac{G_F^2}{12\sqrt{2}\pi^2} M_W^2 
	\frac{M_K}{\Delta M_K} f_K^2 B_K {\rm Im}\biggl\{ 
	(V_{31}^* V_{32}^{})^2 \eta_{K_{33}} F_V^W(3,3) 
	\nonumber \\
	&& ~~~~~
	+ 
	(V_{21}^* V_{22}^{})^2 \eta_{K_{22}} F_V^W(2,2) 
	+ 
	2(V_{31}^* V_{32}^{} V_{21}^* V_{22}^{}) \eta_{K_{32}} 
	F_V^W(3,2)
	\biggr\}, 
\label{epsilon_k}
\eea 
where $f_K$, $B_K$ and $\eta_{K_{ij}}$ represent the 
decay constant, the bag parameter and the QCD correction 
factors, respectively. 

Experimentally, both $\xd$  and $\ek$ parameters have been 
measured as~\cite{PDG}
\bsub
\bea
\xd   &=& 0.73 \pm 0.05, \\
|\ek| &=& (2.23\pm 0.013) \times 10^{-3}. 
\eea
\label{eq:exp_values_bbkk}
\esub
\\
In theoretical estimation of these quantities, 
non-negligible uncertainties come from the evaluations of the 
QCD correction factors and the hadronic matrix elements.  
In our analysis, we adopt the following values:
\bea
\eta_B = 0.55 \pm 0.01~\cite{bb_qcd_buras},
~~~
\sqrt{B_B}f_B = (220 \pm 40)~{\rm MeV}~\cite{fb_const},  
\label{eq:uncertainty_bb}
\eea
for the $\xd$ parameter, and 
\bea
\left.
\begin{array}{lcl}
\eta_{K_{33}} &=& 0.57 \pm 0.01 \\
\eta_{K_{22}} &=& 1.38 \pm 0.20 \\
\eta_{K_{32}} &=& 0.47 \pm 0.04 
\end{array}
\right \}~\cite{bb_qcd_buras,qcd_kk}, 
~~~
B_K = 0.75 \pm 0.15~\cite{recent_estimation}. 
\label{eq:uncertainty_kk}
\eea
for the $\ek$ parameter. 

Next, we consider the new physics contributions to these 
quantities, Br($\kplusdecay$) (\ref{br_kdecay}), 
$\xd$ (\ref{xd_bb}), and $\ek$ (\ref{epsilon_k}). 
In those class of new physics models which have the same 
FCNC structure with that of the SM, the effective Lagrangians 
can be obtained by replacing $D_W^{}(i)$ 
with $D^{\rm new}(i)$ in (\ref{lagrangian_kdecay}),  
and $F_V^W(i,j)$ with $F_V^{\rm new}(i,j)$ in 
(\ref{xd_bb}) and (\ref{epsilon_k}). 
 Then, the effective Lagrangians of these processes in the 
new physics sector should have the following forms;
\bsub
\bea
{\cal L}^{K^+}_{\rm new} &=& \frac{G_F}{\sqrt{2}} 
	\frac{2\alpha(m_Z^{})}{\pi} 
	\frac{1}{\sin^2 \theta_W} 
	\ov{\nu}\gamma^\mu P_L \nu\ 
	V_{32}^* V_{31}^{}\ 
	\ov{s}\gamma_\mu P_L d\ A^{\rm new}, \\
\vsk{0.2}
{\cal L}^{\Delta B =2}_{\rm new} 
	&=& \frac{G_F^2 M_W^2}{4\pi^2}~ 
	\ov{d}\gamma^\mu P_L b~ \ov{d}\gamma_\mu P_L b~
	(V_{31}^* V_{33}^{})^2~ 
	B^{\rm new}, 
\label{eq:del_b=2}  \\
\vsk{0.2}
{\cal L}^{\Delta S =2}_{\rm new} 
	&=& \frac{G_F^2 M_W^2}{4\pi^2}~ 
	\ov{d}\gamma^\mu P_L s~ \ov{d}\gamma_\mu P_L s~
	(V_{31}^* V_{32}^{})^2~ 
	B^{\rm new}. 
\label{eq:del_s=2}
\eea
\esub
It should be noticed that the new physics contributions to 
the $\Delta B=2$ (\ref{eq:del_b=2}) and 
the $\Delta S=2$ (\ref{eq:del_s=2}) 
processes are expressed by the same quantity $B^{\rm new}$. 

There are two cases in which the effective Lagrangians can be 
given by the above forms. 
First, if the contributions from the first two generations 
do not differ much, \ie, 
\bsub
\bea
D^{\rm new}(2) &\approx& D^{\rm new}(1), \\
F_V^{\rm new}(i,1) &\approx&  F_V^{\rm new}(i,2), 
\eea
\esub
the net contributions from the new physics are written by 
using the unitarity of the CKM matrix as; 
\bsub
\bea
\sum_i V_{i2}^* V_{i1}^{} D^{\rm new}(i) &\approx& 
V_{32}^* V_{31}^{} \bigl \{ D^{\rm new}(3) - 
D^{\rm new}(1) \bigr \}, \\
\vsk{0.2}
\sum_{i,j} V_{i1}^* V_{ik}^{} V_{j1}^* V_{jk}^{} F_V^{\rm new}(i,j) 
&\approx& 
(V_{31}^* V_{3k}^{})^2 \bigl\{ F_V^{\rm new}(3,3) 
+ F_V^{\rm new}(1,1) 
\nonumber \\
\vsk{0.1}
&&~~~~~~~~~~~~
- F_V^{\rm new}(3,1) - F_V^{\rm new}(1,3)
\bigr\}, 
\eea
\label{eq:unitarity_cancellation}
\esub
for $k=2,3$. 
We can now define the parameters $A^{\rm new}$ and 
$B^{\rm new}$ as
\bsub
\bea
A^{\rm new} &\equiv& D^{\rm new}(3) - D^{\rm new}(1), \\
B^{\rm new} &\equiv& F_V^{\rm new}(3,3) + F_V^{\rm new}(1,1) 
	- F_V^{\rm new}(3,1) - F_V^{\rm new}(1,3).
\eea
\esub
Second, if the contributions from both the first two 
generations are negligible as compared with those of the 
3rd generation, 
\ie, 
\bsub
\bea
D^{\rm new}(3) &\gg& D^{\rm new}(1), D^{\rm new}(2), \\
F_V^{\rm new}(3,3) &\gg& 
F_V^{\rm new}(1,j), F_V^{\rm new}(2,j), 
F_V^{\rm new}(3,1), F_V^{\rm new}(3,2), 
\eea
\label{eq:yukawa_cancellation}
\esub
the parameters $A^{\rm new}$ and $B^{\rm new}$ become 
\bsub
\bea
A^{\rm new} &=& D^{\rm new}(3), \\
B^{\rm new} &=& F_V^{\rm new}(3,3). 
\eea
\esub

Now, the effects of the new physics contributions to these 
processes can be evaluated by the following ratios 
\bsub
\bea
R_1 &=& \frac{F_V^W(3,3) + B^{\rm new}}{F_V^W(3,3)}, 
\label{r1}
\\[3mm]
R_2 &=& \frac{D_W^{}(3) + A^{\rm new}}{D_W^{}(3)}. 
\eea
\esub
Once a model of new physics is specified, we can quantitatively 
estimate the new contributions in terms of $R_1$ and $R_2$.  
The parameter $R_1$ in (\ref{r1}) has been introduced in 
\cite{bb_mixing} to measure the MSSM contributions to 
the $\xd$  and $\ek$ parameters. 
Both $R_1$ and $R_2$ parameters converge to unity as the new 
physics contributions are negligible, 
\beq
R_1, R_2 \longto 1~~~~{\rm for}~~~
A^{\rm new}, B^{\rm new} \longto 0.
\eeq

In the following, we consider the cases where the net 
contributions from new physics do not exceed those of the SM: 
$A^{\rm new} <  |D_W^{}(3)|$ and $B^{\rm new} < |F_V^W(3,3)|$.  
Then we study constraints on $R_1,R_2$ from experimental 
results in the region of $0 <  R_1,R_2 < 2$. 
For instance, in the MSSM and the THDM, predictions are found 
in the region $0 <  R_1,R_2 < 2$ as shown in Sec. 4.  
\section{Constraints on the new physics contributions to FCNC processes}
\clean
If new physics contributions to $\xd,\ek$ and Br($\kplusdecay$) 
are sizable, the effects can be detected as deviations 
of $R_1$ and $R_2$ from unity. 
In practice, experimentally measurable quantities are 
products of the $R_1$ or $R_2$ by the CKM matrix 
elements. 
In the standard parametrization of the CKM matrix, 
the uncertainty in the CP-violating phase $\delta$ 
dominates that of the CKM matrix elements~\cite{PDG}. 
Hence, together with $R_1$ and $R_2$, we allow $\cosd$ to be 
fitted by the measurements of $\xd,\ek$ and Br($\kplusdecay$). 
For this reason, constraints on $R_1$ and $R_2$ are correlated 
through $\cosd$. 

We perform the $\chi^2$-fit for two parameters $R_1$ and 
$\cosd$ by using experimental data of $\xd$ and 
$\ek$. 
In our fit, we take into account of the theoretical uncertainties 
which are given in (\ref{eq:uncertainty_bb}), 
(\ref{eq:uncertainty_kk}) and 
\bea
\left.
\begin{array}{rcl}
|V_{12}| &=& 0.2205 
 \\
|V_{23}| &=& 0.041 \pm 0.003 \\
|V_{13}/V_{23}| &=& 0.08 \pm 0.02
\end{array}
\right \}~\cite{PDG}, 
~~~
m_t^{}= 175.6 \pm 5.5~{\rm GeV}~\cite{mt96}, 
\eea
where the error of $|V_{12}|$ can be safely neglected. 
We find 
\bea
\begin{array}{l}
	\left.
	\begin{array}{lcl}
	\cosd &=& 0.36 \pm 0.83\\
	\vsk{0.2}
	R_1  &=& 0.93 \pm 0.75
	\end{array}
	\right \}~~~~ 
	\rho_{\rm corr} = 0.90.\\
\vsk{0.3}
\end{array}
\label{eq:r1_cosd}
\eea
Because of the strong positive correlation between 
the errors, only the following combination is effectively 
constrained; 
\beq
R_1 = 0.61 + 0.89 \cosd \pm 0.33.
\eeq
We show the 1-$\sigma$ (39\%) allowed region 
of $\cosd$ and $R_1$ in Fig.~\ref{bbkk39cl}. 
In the figure, there is small region which corresponds to 
$1 \le \cosd$ where the flavor mixing does not obey the CKM 
mechanism. 
\figbbkkonesigma  
The range of $\cosd$ along the $R_1=1$ line is the allowed 
region of $\cosd$ in the SM: $0.08\, \lsim \cosd\, \lsim\, 0.78$. 
We can read off from Fig.~\ref{bbkk39cl} that the current experimental 
data of $\xd$ and $\ek$ parameters constrain the new physics 
contributions within $0.18\, \lsim\, R_1\, \lsim\, 1.68$. 

Next we examine the constraint on $R_2$. 
Although the recent observation of one candidate event is unsuitable 
to include in the actual fit, we can expect that the data will be 
improved in the near future. 
In the following, we adopt the central value of the SM prediction 
as the mean value of Br($\kplusdecay$) and study consequences of 
improved measurements. 
With several more events, the branching fraction 
can be measured as 
${\rm Br}(\kplusdecay) = (0.9 \pm 0.4) \times 10^{-10}$.
Then the combined result with $\xd$ and $\ek$ parameters can be 
found as 
\bea
\begin{array}{l} \left.
	\begin{array}{lcl}
	\cosd &=& 0.36 \pm 0.83\\
	\vsk{0.2}
	R_1  &=& 0.93 \pm 0.75 \\
	\vsk{0.2}
	R_2  &=& 1.14 \pm 0.53
	\end{array}
	\right \}~~~~ 
	\rho_{\rm corr} = \left ( 
	\begin{array}{rrr}
	1 &  0.90 & 0.68  \\
	  &     1 & 0.62   \\
	  &        &   1  
		\end{array} \right ).\\
\vsk{0.3}
\end{array}
\label{eq:r1_r2}
\eea
In Fig.~\ref{sum_r1r2}, 
the results are shown on the $R_1$-$R_2$ plane for 
three values of $\cosd$; 
$\cosd = 0.36$ (mean value), $-0.47$ (${\rm mean\, value}-1\sigma$)
and 1.19 (${\rm mean\, value}+ 1\sigma$). 
\figsumr1r2 
Using this result, we can discuss about constraints on the new 
physics contributions to these processes on the $R_1$-$R_2$ plane 
for a given value of $\cosd$. 
\section{Constraints on MSSM and THDM contributions to 
the FCNC processes}
\clean
Our assumptions on the properties of new physics for FCNC 
processes in $B$ or $K$ meson systems are naturally satisfied 
in both the MSSM and the THDM. 
Predictions on those processes in the contexts of the MSSM and 
the THDM have been studied in \cite{bb_mixing,BB_SUGRA, 
BB_goto,BB_THDM} for $\bbbar,\kkbar$ mixings, and 
\cite{susy_process,thdm_process,susy_buras} 
for $\kplusdecay$ process.  
In this section, we evaluate the $R_1,R_2$ parameters 
in both models and find constraints on them 
from the result in the previous section. 

In the MSSM based on $N=1$ supergravity~\cite{SUGRA}, 
degeneracy of squark masses between the first two 
generations holds in good approximation. 
The interaction vertices among down-type quarks ($d_i$), 
up-type squarks ($\tilde{u}_j$) and charginos ($\chg$) 
are proportional to the CKM matrix elements $V_{ij}$.
Since the top-quark mass could induce the large left-right 
mixing in the $t$-squark sector, one of the $t$-squarks in the 
mass eigenstates can become lighter than the other squarks. 
Presence of such a light $t$-squark weakens the unitarity 
cancellation among the chargino--$u_j$-squark exchange diagrams. 
Therefore the sizable new contributions to the processes may 
arise from the lighter $t$-squark and chargino exchange 
diagram. 

The MSSM has the physical charged Higgs boson as a consequence 
of the supersymmetric extension of the Higgs sector. 
The interactions among the charged Higgs boson and 
quarks are the same with those of the type 
II-THDM~\cite{higgs_hunter}.  
The charged Higgs boson interacts with $d_i$ and 
$u_j$-quarks through the Yukawa interactions which are 
proportional to the corresponding quark masses. 
As a result, the charged Higgs boson contributions to the FCNC 
processes are dominated by its interaction with the top-quark. 

There are other sources of FCNC in the MSSM---the 
interactions among $d_i$-quark, down-type squarks  
and neutralinos or the gluino. 
For $\tan\beta \sim O(1)$, where $\tan\beta$ is the ratio of 
the vacuum expectation values of two Higgs fields, 
the left-right mixing in the down-type squark sector is not 
so large because of the smallness of the down-type quark mass. 
Furthermore, it has been studied that these diagrams 
do not give sizable contributions to the FCNC processes 
for $\tan\beta\, \lsim\, 10$~\cite{BB_goto,susy_buras}.  
Hence we study in the region $\tan\beta\, \lsim\, 10$ 
where their contributions are overwhelmed by the 
$t$-squark--chargino and the charged Higgs boson--top-quark 
contributions. 

The expressions for $R_1$ in the MSSM and the THDM 
can be found in \cite{bb_mixing}. 
The MSSM contribution to the decay process 
$\kplusdecay$ is expressed by using $D^{\rm new}$ 
as follows 
\beq
D^{\rm new}(i) = \sum_{m,n,k,\alpha,\beta}
	D_C^{}(i,m,n;\ell,k;\alpha,\beta) 
	+ 	D_H^{}(i,\ell),  
\eeq
where $D_C^{}(i,m,n;\ell,k;\alpha,\beta)$  and $D_H^{}(i,\ell)$
represent the chargino and the charged Higgs boson contributions, 
respectively. 
The chargino contribution $D_C$ is given by 
\bea
D_C^{}(i,m,n;\ell,k;\alpha,\beta) 
    &=& D_C^{(1)} + D_C^{(2)} + D_C^{(3)} + D_C^{(4)}, 
\eea
and 
\bsub
\bea
D_C^{(1)} &=& 
	-\frac{1}{4}
	\biggl( -\frac{1}{2} + \frac{1}{3} \sin^2 \theta_W \biggr)
	|F_{im}^\alpha|^2 f_1(\sa, \rim), \label{mssm_loop_a}\\
\vsk{0.2}
D_C^{(2)} &=& 
	-\frac{1}{4} F_{im}^{\alpha *} F_{im}^\beta \biggl\{
	B_{\alpha\beta}^L f_2( \sa, \sb, \rim ) + 
	B_{\alpha\beta}^R f_3( \sa, \sb, \rim ) \biggr\}, \\
\vsk{0.2}
D_C^{(3)} &=& 
	-\frac{1}{4} F_{im}^{\alpha *} F_{in}^\alpha 
	D_{mn}^{i} f_4(\rim, \rin, \sa), \\
\vsk{0.2}
D_C^{(4)} &=& 
	\frac{1}{16} F_{im}^\alpha F_{im}^{\beta *}
	G_{\ell k}^{\alpha *} G_{\ell k}^\beta 
	Y_1(\sa, \sb, \rim, \tlk), 
\label{mssm_loop_d}
\eea
\label{eq:mssm_loop_full}
\esub
where the indices $(i,\ell)$ denote the squark and slepton 
generations while $(m,n,k)$ represent two squarks or sleptons 
for each generation. 
The indices $(\alpha,\beta)$ stand for the two charginos. 
The terms $\rim, \tlk$ and $\sa$ are defined by 
\bea
\begin{array}{l}
\disp{ 
r_{11} = r_{21} = \frac{m_{{\tilde u}L}^2}{M_W^2},} 
\hspace{1cm}
\disp{ 
r_{12} = r_{22} = \frac{m_{\tilde{u}R}^2}{M_W^2},} 
\hspace{1cm}
\disp{ 
r_{3k} = \frac{m_{\tilde{t}k}^2}{M_W^2}, } \\
\disp{ 
t_{11} = t_{21} = t_{31} = \frac{m_{\tilde{e}L}^2}{M_W^2}, }
\hspace{1cm}
\disp{ 
t_{12} = t_{22} = t_{32} = \frac{m_{\tilde{e}R}^2}{M_W^2},} \\
\disp{ \sa = \frac{m_{\chg \alpha}^2}{M_W^2} }. 
\end{array}
\eea
The coupling constants 
$F_{im}^\alpha, B_{\alpha\beta}^{L({\rm or~}R)}, 
D_{mn}^{i}$ and $G_{\ell k}^\alpha$, and 
the loop functions $f_1 \sim f_4$ and $Y_1$ 
are explicitly shown in Appendices A and B. 
By using the unitarity of the CKM matrix and the 
degeneracy of the squark masses between the first two 
generations, we obtain 
\beq
V_{i2}^* V_{i1}^{}
D_C^{}(i,m,n;\ell,k;\alpha,\beta) 
= V_{32}^* V_{31}^{} \biggl\{
D_C^{}(3,m,n;\ell,k;\alpha,\beta) 
- D_C^{}(1,m,n;\ell,k;\alpha,\beta)\biggr\}, 
\eeq
and the chargino contribution $A^{\rm new} \equiv A_C$ 
is given by 
\beq
A_C\equiv \sum_{m,n,k,\alpha,\beta}\biggl\{ 
D_C(3,m,n;\ell,k;\alpha,\beta) 
	- D_C(1,m,n;\ell,k;\alpha,\beta)\biggr\}. 
\label{eq:a_c}
\eeq

\figmssm
The charged Higgs boson contribution $D_H(i,\ell)$ 
is given by  
\beq
D_H (i,\ell) = D_{HZ}(i) + D_{HH}(i,\ell) + D_{HW}(i,\ell), 
\eeq
and 
\bsub
\bea
D_{HZ}(i) 
       &=&	-\frac{1}{8} \exi \cot^2\beta \biggl[ 
		\frac{\zi}{(\zi-1)^2}\ln \zi - \frac{\zi}{\zi-1}
		\biggr], \\
\vsk{0.2} 
D_{HH}(i,\ell) &=& 
	\frac{1}{16} \exi \ztild_\ell 
	Y_1(\xh, \xh, \exi, \ztild_\ell), \\
\vsk{0.2} 
D_{HW}(i,\ell) &=& 
	\frac{\sqrt{\exi \ztild_\ell}}{2} Y_2(\xh, 1, \exi, \ztild_\ell) 
	   +\frac{1}{8}	\exi \ztild_\ell Y_1(\xh, 1, \exi, \ztild_\ell),  
\\
\vsk{0.1}
&&~~~~~
\xh = \frac{m_H^2}{m_W^2},~~~~
\zi = \frac{m_{ui}^2}{m_H^2},~~~~
\ztild_\ell = \frac{m_{e\ell}^2}{m_H^2}, 
\eea
\esub
where the indices $(i,\ell)$ correspond to the 
quark and lepton generations, respectively. 
$m_H^{}$ being the charged Higgs boson mass and $\beta$ is 
defined as $\tan\beta \equiv v_2/v_1$, 
$v_1$ and $v_2$ are the vacuum expectation values of the 
Higgs fields of the hyper-charge $Y=-1/2$ and 
$Y=+1/2$, respectively. 
The loop function $Y_2$ is given in Appendix B. 
Due to the smallness of the Yukawa couplings for light quarks,  
the top-quark loop functions $(i=3)$ give dominant contributions. 
Then we can write the charged Higgs contribution as 
\beq
A_H \equiv D_H(3,\ell). 
\label{eq:a_h}
\eeq
From (\ref{eq:a_c}) and (\ref{eq:a_h}), $R_2$ in the MSSM 
is defined as 
\beq
R_2 \equiv \frac{D_W^{}(3) + A_C + A_H}{D_W^{}(3)}. 
\label{mssm_r2}
\eeq
\figthdm 

The MSSM has several unknown parameters. 
In order to reduce the number of input parameters in 
numerical study, we express the soft SUSY breaking scalar masses 
in the squark and the slepton sectors by a common mass parameter 
$m_0^{}$. 
Also taking the scalar trilinear coupling $A_f^{}$ 
($f$ denotes squarks or sleptons) 
as $A_f^{} = m_0^{}$, the MSSM contributions can be evaluated by 
using four parameters, $m_0^{}, \tan\beta$, the higgsino mass term 
$\mu$ and the SU(2) gaugino mass term $m_2$. 
In our study, these parameters are taken to be real. 
In Fig.~\ref{r2_mssm}, we show the 
MSSM contributions to $R_1,R_2$ parameters with 
the constraints on these parameters for $\cosd = 0.36$. 
The numerical study was performed in the range of 
$100~\gev < m_0 < 1~{\rm TeV}, |\mu| < 200~\gev$ and 
$m_2 = 200~\gev$ for $\tan\beta = 2$ and $8$. 
We fixed the charged Higgs boson mass at $m_H = 200~{\rm GeV}$. 
This is the reason why the MSSM contributions do not converge 
to $R_1 = 1$ in Fig.~\ref{r2_mssm}. 
We take into account the 
recent estimation of lower mass limits for lighter $t$-squark 
and lighter chargino~\cite{SUSY_mass_limit}: 
$80~{\rm GeV} \le m_{\tilde{t}1}$ and 
$91~{\rm GeV} \le m_{\chg 1}$. 
The MSSM contribution to $R_1$ interferes with that of the SM 
constructively~\cite{bb_mixing,BB_goto,kurimoto}. 
On the other hand, the contribution to $R_2$ interferes with 
that of the SM both constructively and destructively. 

The THDM contribution to $R_2$ is given by setting 
$D_C = 0$ in (\ref{mssm_r2}): 
\beq
R_2 \equiv \frac{D_W^{}(3) + A_H}{D_W^{}(3)}. 
\eeq
We show in Fig.~\ref{r2_thdm} the charged Higgs contribution 
to $R_1,R_2$ parameters for $\tan \beta = 2$ and $\cosd = 0.36$. 
Contrary to the case of the MSSM, the THDM contribution 
constructively interferes with the SM contribution for 
both $R_1$ and $R_2$. 
Here we show the case of $\tan \beta = 2$ only. 
The Yukawa interaction between the top-quark and the 
charged Higgs boson is proportional to 
$1/\tan^2\beta$. 
Thus constraints on the THDM contribution to 
these quantities are weakened together with the increase of $\tan\beta$. 

\section{Summary}
We have studied impacts on searching for signatures of 
new physics beyond the SM from some FCNC processes -- 
$\bbbar, \kkbar$ mixings and the rare decay $\kplusdecay$.  
For a certain class of models of new physics, 
we showed the extra contributions to the FCNC processes can be 
parametrized by its ratio to the SM contribution with the 
common CKM matrix elements. 
Two parameters $R_1$ and $R_2$ were introduced to estimate 
the new physics contributions to $\bbbar, \kkbar$ mixings 
and $\kplusdecay$ decay, respectively. 
Then the new physics contributions are evaluated 
from experimental data by using these parameters 
and $\cosd$. 

Taking account of both experimental and theoretical 
uncertainties for the $\bbbar$ and $\kkbar$ mixings, 
constraints on the new physics contribution to $R_1$ and 
$\cosd$ were shown: the allowed range of $R_1$ is 
$0.18\, \lsim\, R_1\, \lsim\, 1.68$. 
With the assumption that the future data of Br($\kplusdecay$) 
will be close to the SM prediction, 
constraints on $\cosd,R_1$ and $R_2$ were found.  
The results were applied to the MSSM and the THDM 
contributions to those processes. 
Our study will become useful if the measured value of 
Br($\kplusdecay$) is close to the SM prediction. 
Then, we may expect to obtain the constraints on the new physics 
parameters through $R_1$ and $R_2$. 

\section*{Acknowledgment}
The author thanks to K. Hagiwara, Y. Okada and Y. Shimizu for 
discussions and comments. 
This work is supported in part by Grant-in-Aid for Scientific 
Research from the Ministry of Education, Science and Culture 
of Japan.

\vsp{1}
{\bf Note added: } 
While we were preparing this paper, we found ref.~\cite{susy_buras},  
where a parametrization of new physics contributions to 
$\kplusdecay$ is proposed and consequences of the SUSY-SM are studied. 
Their parametrization is similar to ours besides that 
they define the parameter $R_2$ (denoted as $r_K$ in their 
paper) as a complex parameter. 
Our result of the MSSM contributions to the decay process 
is consistent with theirs. 
Correlation between the MSSM contributions to 
$\kplusdecay$ and the $\xd$, $\ek$ parameters are not discussed in 
their paper.

\section*{Appendix A: Masses and coupling constants in the MSSM}
\renewcommand{\theequation}{A. \arabic{equation}}
\clean
In this appendix, we give the explicit forms of coupling 
constants in (\ref{eq:mssm_loop_full}). 
We first introduce the squark, slepton and chargino masses. 
The squark masses in the first and the second generations 
are given by
\bea
\begin{array}{l}
\disp{
m_{\tilde{u}L}^2 = m_{\tilde{c}L}^2 =
m_Q^2 + \cos 2\beta (\frac{1}{2}-\frac{2}{3} \sn2w) m_Z^2, }
\\[3mm]
\disp{
m_{\tilde {u}R}^2 = m_{\tilde {c}R}^2 =
m_U^2 + \frac{2}{3}\cos 2 \beta \sn2w m_Z^2, }
\end{array}
\eea
where the corresponding quark masses can be safely neglected. 
The parameters 
$m_Q$ and $m_U$ are the soft SUSY breaking squark masses  
for the SU(2) doublet and the singlet, respectively. 
The angle $\beta$ is defined by $\tan\beta = v_2/v_1$, 
where $v_1,v_2$ are the vacuum expectation values of the 
Higgs doublets. 
The squared mass matrix for the $t$-squark is given by 
\beq
M_{\tilde t}^2 = \left(
\begin{array}{cc}
m_{\tilde{u}L}^2 + m_t^2 & 
-m_t (\mu\cot\beta + A_t) \\ 
\noalign{\vskip0.2cm}
-m_t (\mu\cot\beta + A_t) &  
m_{\tilde{u}R}^2 + m_t^2   
\end{array}
\right), 
\eeq
where the dimensionful parameter $A_t$ and $\mu$ denote the 
scalar trilinear coupling and the higgsino mass term, respectively. 
The mass matrix $M^2_{\tilde{t}}$ can be diagonalized by 
using the unitary matrix $S_t$, 
\begin{eqnarray}
S_t M^2_{\tilde{t}} S_t^\dagger 
      &=& {\rm diag}(m_{\tilde {t}1}^2, 
	m_{\tilde {t}2}^2)~~~~~
	( m_{\tilde {t}1}^2 < m_{\tilde {t}2}^2 ). 
\end{eqnarray}

The charged slepton masses are given by 
\bea
\begin{array}{l}
\disp{
m_{\tilde{e}L}^2 = m_{\tilde{\mu}L}^2 = 
m_{\tilde{\tau}L}^2 =
m_L^2 + \cos 2\beta (-\frac{1}{2}+ \sn2w) m_Z^2, }
\\[3mm]
\disp{
m_{\tilde {e}R}^2 = m_{\tilde {\mu}R}^2 = 
m_{\tilde {\tau}R}^2 = 
m_E^2 - \cos 2 \beta \sn2w m_Z^2, }
\end{array}
\eea
where $m_L^{}$ and $m_E^{}$ represent the soft SUSY breaking 
slepton masses for the SU(2) doublet and the singlet, 
respectively. 
We neglected the corresponding charged lepton masses. 

The chargino mass matrix is given by 
\begin{equation}
M^- = \left(
\begin{array}{cc}
	m_2 & \sqrt{2}M_W \cos\beta \\
	\sqrt{2}M_W \sin\beta & \mu
\end{array}
\right), 
\end{equation}
where $m_2$ is the SU(2) gaugino mass. 
We can obtain the mass eigenstates by 
using two unitary matrices $C_R^{}$ and $C_L^{}$; 
\begin{equation}
C_R^\dagger M^- C_L = {\rm diag}(\tilde{m}_{\omega 1}, 
	\tilde{m}_{\omega 2})\ \ \ 
	(\tilde{m}_{\omega 1} < \tilde{m}_{\omega 2}). 
\end{equation}

The couplings constants $F_{i j}^\alpha$, $B_{\alpha \beta}^k$ 
($k=L,R$), $D_{\ell m}^i$ and  $G_{i j}^\alpha$ 
in (\ref{eq:mssm_loop_full}) are given as; 
\bea
\left. 
\begin{array}{l}
F_{11}^\alpha = F_{21}^\alpha = \sqrt{2} C_{R1\alpha}^*,  
~~~~F_{12}^\alpha = F_{22}^\alpha = 0, 
\\[3mm]
\disp{
F_{3j}^\alpha = \sqrt{2} C_{R1\alpha}^* S_{tj1}^{} - 
\frac{m_t^{}}{M_W \sin\beta} C_{R2\alpha}^* S_{tj2}^{}}
\end{array}
\right \}, 
\eea
\bea
\begin{array}{l}
\disp{
B_{\alpha\beta}^k = 
-C_{k1\alpha}^*C_{k1\beta}^{}-\frac{1}{2}C_{k2\alpha}^*C_{k2\beta}^{}
+ \delta_{\alpha\beta}\sn2w, }
\end{array}
\eea
\bea 
D_{\ell m}^{3} = \biggl( \frac{1}{2} - \frac{2}{3}\sn2w \biggl)
	S_{t\ell 1} S_{t m1}^* 
	- \frac{2}{3}\sn2w S_{t\ell 2} S_{t m2}^*, 
\label{eq:coupl_d}
\eea
\bea
\left.
\begin{array}{l}
G_{11}^\alpha = G_{21}^\alpha = G_{31}^\alpha 
= \sqrt{2} C_{L1\alpha}^*, \\ [3mm]
G_{12}^\alpha = G_{22}^\alpha = G_{32}^\alpha 
= 0
\end{array}
\right\}, 
\eea
where the expressions for the first two generation 
of squarks in (\ref{eq:coupl_d}) can be obtained 
by replacing $S_t$ with the unit matrix.

\vsp{1}

\section*{Appendix B: Loop functions} 
\renewcommand{\theequation}{B. \arabic{equation}}
\clean

The loop function $f_1 \sim f_4$ in 
(\ref{mssm_loop_a})$\sim$(\ref{mssm_loop_d}) are 
given as;
\bea
f_1(x,y) &=& \frac{1}{4} + \frac{1}{2} \frac{x}{x-y} - \frac{1}{2}
	\biggl\{ \ln y + \biggl( \frac{x}{x-y} \biggr)^2 
	\biggl( \ln x - \ln y \biggr) \biggr\}, 
\\ [3mm]
f_2(\exi, \xj, y) &=& -\sqrt{\exi \xj} \biggl\{ 
	\frac{\exi \ln \exi}{ (\exi - \xj)(\exi - y) }
	+ 
	\frac{\xj \ln \xj}{ (\xj - \exi)(\xj - y) } \nonumber \\[3mm]
	&&\hspace{4cm}
	+ 
	\frac{y \ln y}{ (y - \exi)(y - \xj) }
	\biggr\}, 
\\ [3mm]
f_2(x, x, y) &=& -\frac{x}{y - x}\biggl\{ 
	\frac{y}{y - x} \biggl( \ln y - \ln x \biggr) - 1
	\biggr\}, \\ [3mm]
f_3(\exi, \xj, y) &=& \frac{1}{2} \biggl\{ 
	\frac{\exi^2 \ln \exi}{ (\exi-\xj)(\exi-y) }
	+ 
	\frac{\xj^2 \ln \xj}{ (\xj - \exi)(\xj-y) } \nonumber \\[3mm]
	&&\hspace{4cm}
	+ 
	\frac{y^2 \ln y}{ (y - \exi)(y-\xj) } 
	\biggr\} - \frac{1}{4},\\ [3mm]
f_3(x, x, y) &=& \frac{1}{2} \biggl\{ 
	\biggl( \frac{y}{y-x} \biggr)^2 (\ln y - \ln x)
	+ \ln x - \frac{x}{y-x} 
	\biggr\} - \frac{1}{4}, \\ [3mm]
f_4(\exi, \xj, y) &=& f_3(\exi, \xj, y) - \frac{1}{2},\\ [3mm]
f_4(x, x, y) &=& f_3(x, x, y) - \frac{1}{2}. 
\eea

The loop functions $Y_1, Y_2$ which come from the box type diagrams 
are given by; 
\begin{eqnarray}
Y_1(\ra,\rb,\si,\sj) &=&  
  \frac{r_\alpha^2}{(\rb - \ra)(\si - \ra)(\sj - \ra)}\ln \ra  
+ \frac{r_\beta^2}{(\ra - \rb)(\si - \rb)(\sj - \rb)}\ln \rb, 
	\nonumber \\
\vsk{0.2}
& &\hspace{-1.3cm}
+ \frac{s_i^2}{(\ra - \si)(\rb - \si)(\sj - \si)}\ln \si 
+ \frac{s_j^2}{(\ra - \sj)(\rb - \sj)(\si - \sj)}\ln \sj, 
\\[5mm]
Y_1(\ra,\ra,\si,\sj) &=& 
  \frac{\ra(\si + \sj) - 2\si\sj}{(\si - \ra)^2(\sj - \ra)^2}\ \ra \ln \ra 
- \frac{\ra}{(\si - \ra)(\sj - \ra)} \nonumber \\
\vsk{0.2}
& &\ \  
+ \frac{s_i^2}{(\ra - \si)^2(\sj - \si)}\ln \si 
+ \frac{s_j^2}{(\ra - \sj)^2(\si - \sj)}\ln \sj, 
\\[5mm]
Y_1(\ra,\rb,\si,\si) &=& 
  \frac{r_\alpha^2}{(\rb - \ra)(\si - \ra)^2}\ln \ra 
+ \frac{r_\beta^2}{(\ra - \rb)(\si - \rb)^2}\ln \rb
\nonumber \\
\vsk{0.2}
& &
+ \frac{(\ra + \rb)\si - 2\ra \rb}{(\ra - \si)^2(\rb - \si)^2}\ \si \ln \si 
- \frac{\si}{(\ra - \si)(\rb - \si)},
\\[5mm]
Y_1(\ra,\ra,\si,\si) &=& 
-\frac{2\ra \si}{(\si - \ra)^3}\ln \ra 
-\frac{2\ra \si}{(\ra - \si)^3}\ln \si 
-\frac{\ra + \si}{(\ra - \si)^2},
\\[7mm]
Y_2(\ra,\rb,\si,\sj) &=& 
\sqrt{\si \sj} \biggl [ 
  \frac{\ra}{(\rb - \ra)(\si - \ra)(\sj - \ra)}\ln \ra  
+ \frac{\rb}{(\ra - \rb)(\si - \rb)(\sj - \rb)}\ln \rb 
\nonumber \\[4mm]
&&\hspace{-1.5cm}
+ \frac{\si}{(\ra - \si)(\rb - \si)(\sj - \si)}\ln \si 
+ \frac{\sj}{(\ra - \sj)(\rb - \sj)(\si - \sj)}\ln \sj \biggr ],
\\[5mm]
Y_2(\ra,\ra,\si,\sj) &=& \sqrt{\si \sj} \biggl [
  \frac{r_\alpha^2 - \si\sj}{(\si - \ra)^2(\sj - \ra)^2}\ln \ra  
- \frac{1}{(\si - \ra)(\sj - \ra)} \nonumber \\
\vsk{0.2}
& &
+ \frac{\si}{(\ra - \si)^2(\sj - \si)}\ln \si 
+ \frac{\sj}{(\ra - \sj)^2(\si - \sj)}\ln \sj \biggr ],
\\[5mm]
Y_2(\ra,\rb,\si,\si) &=& \si \biggl [
\frac{\ra}{(\rb - \ra)(\si - \ra)^2}\ln \ra 
+ \frac{\rb}{(\ra - \rb)(\si - \rb)^2}\ln \rb 
\nonumber \\
\vsk{0.2}
& &\ \ 
+ \frac{s_i^2 - \ra\rb}{(\ra - \si)^2(\rb - \si)^2}\ln \si  
- \frac{1}{(\ra - \si)(\rb - \si)} \biggr ],
\\[5mm]
Y_2(\ra,\ra,\si,\si) &=& \si \biggl [
-\frac{\ra + \si}{(\si - \ra)^3}\ln \ra 
-\frac{\ra + \si}{(\ra - \si)^3}\ln \si 
-\frac{2}{(\ra - \si)^2} \biggr ].
\end{eqnarray}

\newpage

\end{document}